**On-line abstract:**

The 11-year activity cycle of the Sun is a consequence of a dynamo process occurring beneath its surface. We analyzed photometric data obtained by the CoRoT space mission, showing solar‑like oscillations in the star HD49933, for signatures of stellar magnetic activity. Asteroseismic measurements of global changes in the oscillation frequencies and mode amplitudes reveal a modulation of at least 120 days, with the minimum frequency shift corresponding to maximum amplitude as in the Sun. These observations are evidence of a stellar magnetic activity cycle taking place beneath the surface of HD49933 and provide constraints for stellar dynamo models under conditions different from those of the Sun.

# CoRoT reveals a magnetic activity cycle in a Sun-like star

Rafael A. García,[1*] Savita Mathur,[2] David Salabert,[3,4] Jérôme Ballot,[5] Clara Régulo,[3,4] Travis S. Metcalfe,[2]   Annie Baglin[6]

[1]*Laboratoire AIM, CEA/DSM-CNRS, Université Paris 7 Diderot, IRFU/SAp-SEDI, Centre de Saclay, 91191, Gif-sur-Yvette, France.* [2]*High Altitude Observatory, NCAR, P.O. Box 3000, Boulder, CO 80307, USA.* [3]*Instituto de Astrofísica de Canarias, 38200, La Laguna, Tenerife, Spain.* [4]*Depto. de Astrofísica, Universidad de La Laguna, 38206 La Laguna, Tenerife, Spain.* [5]*Laboratoire d'Astrophysique de Toulouse-Tarbes, Université de Toulouse, CNRS, 31400, Toulouse, France.*[6]*LESIA, UMR8109, Université Pierre et Marie Curie, Université Denis Diderot, Obs. de Paris, 92195 Meudon Cedex, France.*

*To whom correspondence should be addressed. E-mail: Rafael.Garcia@cea.fr

**Brevia Text:**

Our understanding of and capability to predict the 11-year activity cycle of the Sun (1), which is driven by a dynamo process seated at the bottom of the convective zone, is far from being complete, as suggested by the recent unusually long solar minimum of cycle 23 (2, 3). Observations of magnetic activity cycles in other stars can potentially improve our knowledge of dynamo processes because they allow us to sample different physical conditions distinct from those in the Sun. Although many stars exhibit activity cycles and some empirical relations have been found (4), a detailed knowledge of the internal structure and dynamics of the star, in particular the depth of the convective envelope and the degree of differential rotation, is required to improve the constraints on theory. Here we show how asteroseismology has revealed the signature of a magnetic activity cycle in a Sun-like star.

Helio- and asteroseismology are the only tools available that can probe the structure and dynamics of the Sun and stars through the study of acoustic oscillations (5). The oscillation modes are sensitive to variations in the magnetic field, and their characteristics change throughout the activity cycle. In the Sun, for instance, the oscillation frequencies are shifted higher during solar maximum while the amplitudes decrease, showing anti-correlated temporal variations (6-9).

The Convection Rotation and planetary Transits (CoRoT) mission has so far observed 6 main-sequence stars exhibiting solar-like oscillations during more than 130 days each as part of its asteroseismic program (10). One of these stars is HD49933, an F5V star 20% more massive and 34% larger than the Sun. It has been observed twice, for 60 days in 2007 and 137 days in 2008, spanning a total of 400 days. More than 50 individual acoustic modes have been identified (11, 12) and then used to model the stellar interior (e.g. 13). Its rotation period is 3.4 days (8 to 9 times faster than Sun).

To search for the effects of magnetic activity in this star, we measured the variation of the mode amplitude and the frequency shift of the acoustic-mode envelope (7). As observed in the Sun, there is a clear anti-correlated temporal variation between both parameters (Fig. 1), revealing a modulation in the second epoch that seems to indicate a period of at least 120 days related to the internal magnetic activity of HD49933. The observations agree with the scaling proposed by (8), which predicts larger than solar frequency shifts for stars that are hotter and more evolved, in contradiction to the scaling suggested by (9). The long-period variations detected in the luminosity of the star –interpreted as fluctuations in the positions and sizes of starspots– allow us to also study the surface activity (7). Indeed, the starspot signature changes with time showing a quiet period during the first part of the second set of observations (Fig. 1). This confirms the existence of an activity cycle, which seems to be shifted in time compared to the seismic indicators. Finally, medium resolution spectra of the Calcium H and K lines obtained on April 13, 2010 confirm that HD49933 is an active star with a Mount Wilson S-index of 0.3. Asteroseismology has thus revealed a stellar activity cycle analogous to that of the Sun.

References and Notes

1. P. Charbonneau, *Living Reviews in Solar Physics*, **2**, 2 (2005).

2. D. Salabert *et al., Astron. Astrophys.* **504**, L1 (2009).

3. S.C. Tripathy, K. Jain, F. Hill, J.W. Leibacher, Astrophys. J. **711**, L84 (2010).

4. S.L. Baliunas *et al.*, *Astrophys. J.* **438**, 269 (1995).

5. D.O. Gough *et al., Science* **272**, 1296 (1996).

6. S.J. Jiménez-Reyes *et al., Astrophys. J.,* **604**, 969 (2004).

7. See supporting material on Science Online.

8. T.S. Metcalfe, W.A. Dziembowski, P.G. Judge, M. Snow, *MNRAS* **379**, L16 (2007).

9. W.J. Chaplin, Y. Elsworth, G. Houdek, R. New, *MNRAS* **377**, 17 (2007).

10. E. Michel *et al., Science* **322**, 558 (2008).

11. T. Appourchaux, E. Michel, M. Auvergne *et al., Astron. Astrophys.* **488**, 705 (2008).

12. O. Benomar *et al., Astron. Astrophys.* **507**, L13 (2009).

13. L. Piau, S. Turck-Chièze, V. Duez, R.F. Stein*., Astron. Astrophys.* **506**, 175 (2009).

14. The CoRoT space mission has been developed and is operated by CNES, with contributions from Austria, Belgium, Brazil, ESA (RSSD and Science Program), Germany and Spain. This work has been partially supported by the Spanish National Research Plan (grant PNAyA2007-62650) and by the "Programme National de Physique Stellaire" at CEA/Saclay. NCAR is supported by the National Science Foundation.

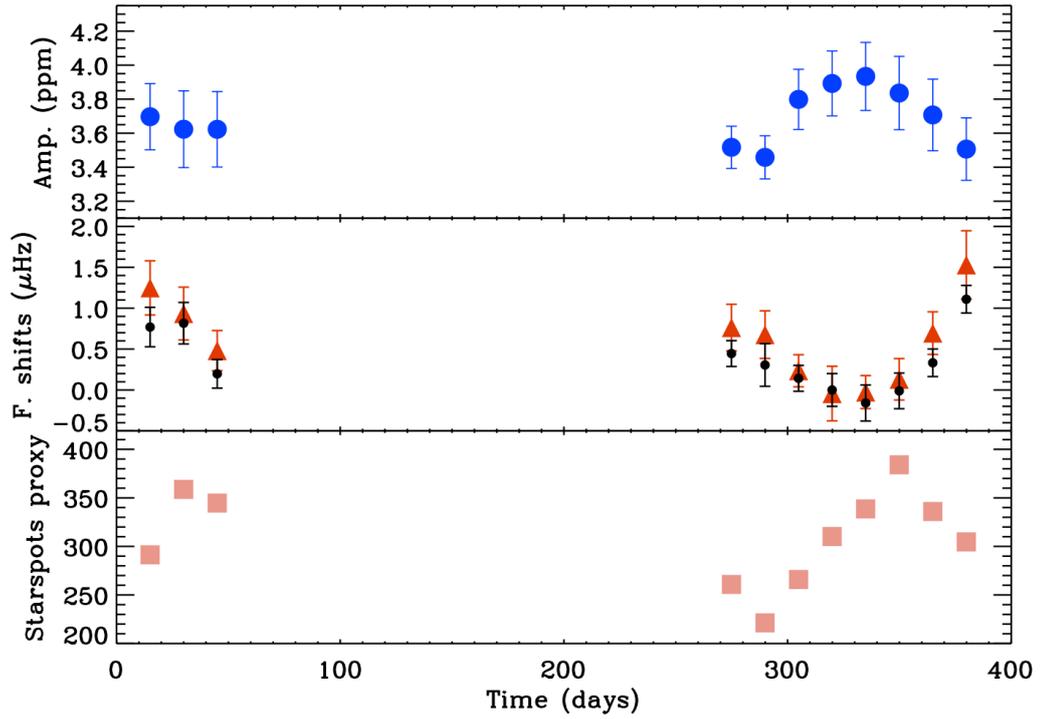

**Fig. 1.** Time evolution –beginning February 6, 2007– of the mode amplitude **(top)**, the frequency shifts using two different methods **(central)**: cross correlations (red triangles) and individual frequency shifts (black circles); and a starspot proxy **(bottom)** built by computing the standard deviation of the light curve (7). All of them are computed using 30-day long subseries shifted every 15 days (50% overlapping). The corresponding 1σ error bars are shown.

# Supporting Online Material for
## CoRoT reveals a magnetic activity cycle in a Sun-like star

Rafael A. García,[1*] Savita Mathur,[2] David Salabert,[3,4] Jérôme Ballot,[5] Clara Régulo,[3,4] Travis S. Metcalfe,[2] Annie Baglin[6]

[1]Laboratoire AIM, CEA/DSM-CNRS, Université Paris 7 Diderot, IRFU/SAp-SEDI, Centre de Saclay, 91191, Gif-sur-Yvette, France. [2]High Altitude Observatory, NCAR, P.O. Box 3000, Boulder, CO 80307, USA. [3]Instituto de Astrofísica de Canarias, 38205, La Laguna, Tenerife, Spain. [4]Departamento de Astrofísica, Universidad de La Laguna, Tenerife, Spain. [5]Laboratoire d'Astrophysique de Toulouse-Tarbes, Université de Toulouse, CNRS, 31400, Toulouse, France.[6]LESIA, UMR8109, Université Pierre et Marie Curie, Université Denis Diderot, Obs. de Paris, 92195 Meudon Cedex, France.

## 1.- Methodology:

CoRoT is a 27 cm afocal telescope, imaging the stellar field onto four CCDs (2048 x 4096 pixels each), two dedicated for exoplanet research and two for asteroseismology. In this work, we have analyzed the so-called "Helreg" (S1) light curves of HD49933 provided by the CoRoT satellite with a regular sampling rate of 32 seconds.

We divided these observations into subseries of 30 days shifted by 15 days and computed the associated power spectra using a standard Fourier Transform algorithm. Then we subtracted the signal from the convective background using a one component standard Harvey model (see a detailed explanation in S2).

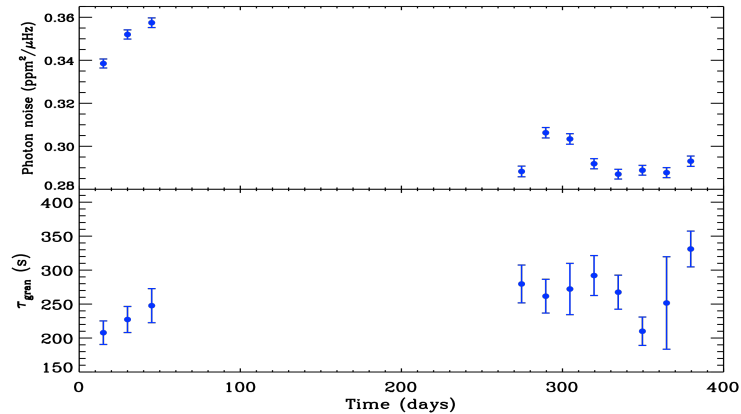

**Figure S1:** Variations with time of the photon noise **(top)** and the characteristic time of the granulation **(bottom).**

Fig. S1 shows the time evolution of the photon noise and the granulation time scale ($\tau_{gran}$). In both cases, there is no correlation with the seismic parameters. This is also the case in the Sun, for which no correlation has been found between the granulation characteristic time and several activity proxies during solar cycle 23 (S3).

Due to the short length of the subseries, the frequency resolution of the power spectrum is low and we started by using the full p-mode envelope instead of analyzing the amplitude and the frequency of each individual mode. Thus, for each subseries the root-mean-square maximum amplitude per radial mode was computed by fitting a Gaussian function to the p-mode region between 1400 to 2500 µHz where this bump is visible (S2).

The frequency shift of acoustic modes has been used several times as a good proxy of the activity cycle (see for example S4, S5). In the present analysis it is calculated by two methods: the cross correlation of the p-mode region of different data segments and by computing the time evolution of the average of the individual p modes.

### 1.1.-Frequency shift compute by cross-correlation methods:

For each subseries, we calculate the cross correlation of the power spectrum (without smoothing the spectrum) with a reference one, taken during the minimum of activity in the region where the individual modes can be identified (1460 to 2100 µHz). As the cross-correlation function is not symmetrical, we compute the third moment of the distribution of points (see Fig. S2) giving its skewness. To obtain the maximum of the cross-correlation function, we fit a Gaussian profile centered at the position given by the skewness to avoid any bias induced by this asymmetry. The frequency shift is then given by the centroid of the fitted Gaussian. The associated error is the 1σ error bar of the fit. This method has been extensively used to study the activity cycle using single-site helioseismic ground-based data (S6). In Fig. S2 we show two examples of the correlation functions found in the present analysis. On the left, the cross-correlation between the first subseries and the reference one is shown, while on the right we plot the cross correlation between the last subseries and the reference one.

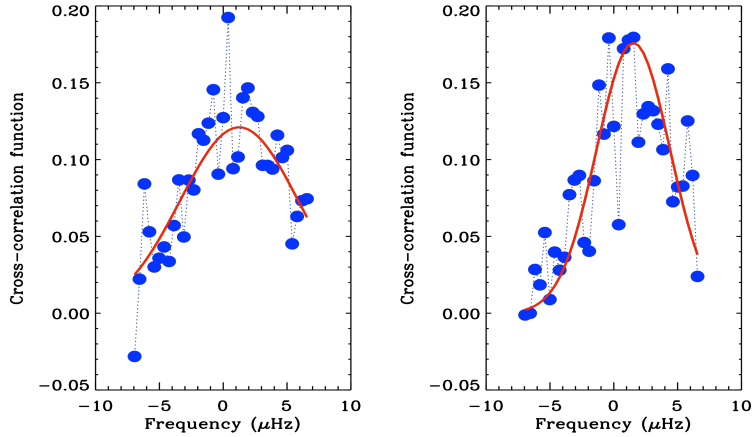

**Figure S2:** Examples of cross-correlation functions used to compute the frequency shift. We have used the 7$^{th}$ subseries as a reference because it corresponds to the minimum activity (see Fig. 1). The continuous red line is the Gaussian fit used to obtain the frequency shift. The left panel shows the cross-correlation between the first subseries and the reference one. The right panel shows the one between the last subseries and the reference one.

### 1.2.-Time evolution of average individual frequency shifts of HD49933

Even with a relatively low signal-to-noise ratio, the individual p-mode parameters can be extracted by fitting each subseries of 30 days using a standard likelihood maximization function (S7). Each mode component was described by a Lorentzian profile. The fitting was performed over frequency windows containing the l = 0, 1, and 2 modes. The initial parameters were taken from (12), the inclination angle and the rotational splitting were fixed to their estimated values (12). The temporal variation of the l = 0 and 1 modes was calculated over the frequency range 1460 - 2070 µHz, the outliers above 5σ were removed. The resulting frequency shifts have the same temporal dependence and amount of change as those obtained with cross-correlation methods (see Fig. 1).

### 2.-Application to the Sun.

In order to test our methodology we have applied it to solar photometric data obtained from the SPM/VIRGO instrument on board the SoHO spacecraft (S8) during solar cycle 23. We first computed the average of the three independent channels (red, blue, and green) and we divided the full time series into subseries of 1 year with 25% overlap from 1996 till 2006. This subseries length was chosen to avoid any periodicity due to the SoHO orbital period and to maintain a proportion of ~10% between the size of the subseries and the length of the activity cycle, which is close to what appears to be the case for HD49933. The frequency range used in the analysis is from 1900 to 5000 µHz for the amplitudes and between 2400 and 3400 µHz for the frequency shifts. Fig. S3 shows the temporal variation of the p-mode amplitude as well as the frequency shift obtained from the cross-correlation method and from the analysis of individual p-mode frequencies. The anti-correlation between the frequency shift and the amplitude variation is apparent. We also show the International Sunspot Number –obtained from the Solar Geophysical Data Centre (*http://www.ngdc.noaa.gov/*)– as a proxy of surface activity. Seismic indicators vary in phase with solar activity proxies, but the degree of correlation differs depending on the phase of the cycle (S9).

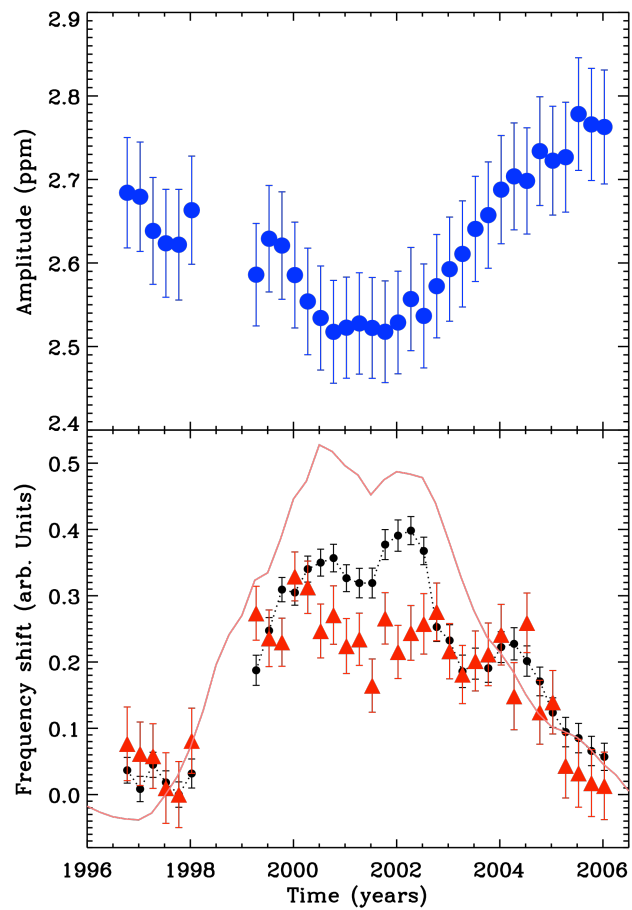

**Figure S3:** Temporal evolution of the p-mode amplitude **(top)** and the frequency shift **(bottom)** using solar SPM/VIRGO data. The red triangles are the frequency shifts computed using the cross-correlation method. The black circles are the frequency shifts computed using individual frequencies of modes l=0 to 2. We have used the 5$^{th}$ series as the reference (with a frequency shift 0). The continuous line is the International Sunspot Number. We have removed the series for which the duty cycle is less than 80% (corresponding to the SoHO vacation period in the second part of 1998 and beginning 1999).

It is important to notice that the frequency shift measured in the Sun using cross-correlation methods is smoother compared to the average frequency shift obtained from the analysis of individual modes. This is a consequence of the high signal-to-noise ratio of solar observations and the different weighted average implicit in each method. In the case of the individual frequency shifts, the modes l=0, 1 and 2 have the same weight while in the cross-correlation methods the modes are weighted by their intrinsic amplitudes (l=1 are higher than l=0 and l=2). Since the frequency shift is not the same for each degree, the global shape is the same but in the details there are some differences.

## 3.- Light Curve of HD49933

A visual inspection of the light curve of HD49933 (see Fig. S4) reveals a modulation of a few days that has been interpreted as the signature of starspots crossing the visible surface of the star. Spot modeling for the first set of observations provides an estimate of the rotation rate of the star, its inclination angle, and the lifetime of the starspots that was established to be around 3.45 days (S10). Moreover, we see quiet periods in which the starspot signature is small and others in which the signature of the rotation is better defined.

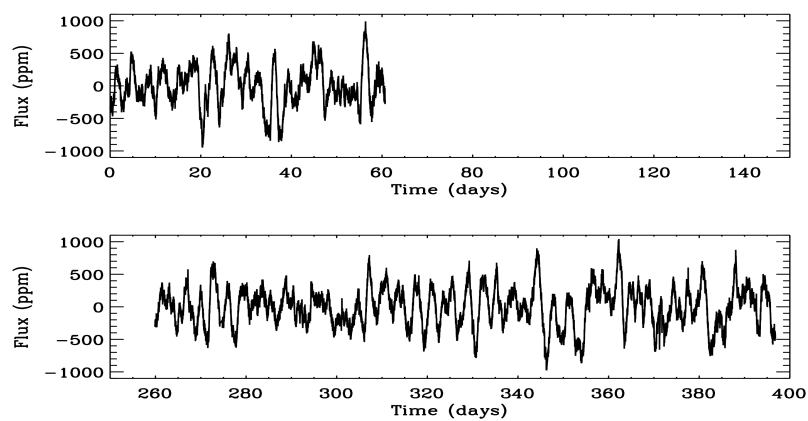

**Figure S4:** Light curve of HD49933 for the first set of measurements **(top)** and the second one **(bottom)**. We have rebinned the light curve to 640s.

To better uncover any signature of an activity cycle in the light curve we have built a starspot proxy by computing the standard deviation of the light curve using the same subseries of 30 days shifted every 15 days. The result is shown in the bottom panel of Fig. 1.

Once again, we see a clear modulation in this activity proxy with a maximum that is shifted around 30 days compared to the seismic indicators. The rising phase takes around 60 days to evolve from the minimum to the maximum. Unfortunately, as it is quite possible that the cycle would not be symmetric (the slope of the rising phase being probably different than the descending phase), we are not able to give the full length of the cycle. Indeed, to correctly interpret this activity proxy –obtained from the light curve– we need to take into account the small inclination angle of the star, $17°^{+7}_{-9}$ (12), and make some assumptions about the active longitudes of the star.

## 4.- Observations of Ca H and K.

Two medium resolution spectra of HD49933 were obtained with the RCSpec instrument on the SMARTS 1.5m telescope at Cerro Tololo on 2010 April 13 (fits files are available as part of the support on-line material). The data were subjected to the usual bias and flat field corrections, and the spectra were extracted and wavelength calibrated using a reference Th-Ar spectrum obtained immediately before the stellar exposures. Following (S11) the calibrated spectra were then integrated in 1.09 Å triangular passbands centered on the cores of the Ca H and K lines and compared to 20 Å continuum passbands from the wings of the lines to generate an S-index on the Mount Wilson system. These spectra confirm that HD49933 is an active star, with a Mount Wilson S-index of 0.302 ± 0.005.

**References and Notes**

S1. M. Auvergne *et al., Astron. Astrophys*. **506**, 411 (2009).

S2. S. Mathur *et al., Astron. Astrophys*. **511**, A46 (2010).

S3. S. Lefebvre *et al.*, *Astron. Astrophys*. **490**, 1143 (2008).

S4. M.F. Woodard, R.W. Noyes, *Nature* **318**, 449 (1985).

S5. S.J. Jiménez-Reyes, C. Régulo, T. Roca Cortés, *Astron Astrophys*. **329**, 1119 (1998).

S6. P.L. Pallé, C. Régulo, T. Roca Cortés, *Astron Astrophys*. **224**, 253 (1989).

S7. T. Appourchaux, L. Gizon, M. C. Rabello Soares, *Astron. Astrophys*. **132**, 107 (1998).

S8. C. Frohlich *et al.*, *Solar Phys.* **162**, 101 (1995).

S9. K. Jain, S. C. Tripathy, F. Hill, *Astrophys. J.* **695**, 1567 (2009).

S10. B. Mosser *et al.*, *Astron. Astrophys.* **506**, 245 (2009).

S11 D.K. Duncan *et al.*, *Astrophys J. Suppl.*, **76**, 383 (1991).